\documentclass[useAMS,usenatbib]{mnras}

\usepackage{graphicx}
\usepackage{float}
\usepackage{hyperref}
\usepackage{color}
\usepackage{amsmath} 
\usepackage{amssymb} 
\usepackage{mathrsfs}

\newcommand{\mum}{${\rm \mu m}$}
\newcommand{\Ledd}{${\lambda_{\rm Edd}}$}
\newcommand{\sampsize}{1620}

\title[An enhanced fraction of starburst at high \Ledd.]{An enhanced fraction of starbursting galaxies among high Eddington ratio AGNs.}

\author[E.~Bernhard et al.]
{\parbox{\textwidth}{E.~Bernhard,$^{1}$\thanks{E-mail: \texttt{ebernhard1@sheffield.ac.uk}}
    J.R.~Mullaney,$^{1}$
    E.~Daddi,$^{2}$
    L.~Ciesla,$^{2}$  and
    C.~Schreiber$^{2,3}$}
\vspace{0.4cm}\\\
\parbox{\textwidth}{
  $^{1}$Department of Physics $\&$ Astronomy, University of Sheffield, Sheffield S3 7RH, UK\\
  $^{2}$Laboratoire AIM-Paris-Saclay, CEA/DSM/Irfu - CNRS - Universit\'e Paris Diderot, CEA-Saclay, pt courrier 131, F-91191 Gif-sur-Yvette, France\\
  $^{3}$Leiden Observatory, University of Leiden, PO Box 9513, 2300 RA Leiden, The Netherlands}}

\begin{document}

\date{Accepted 2016 April 21; Received 2016 April 7; in original form 2016 February 5}

\maketitle

\begin{abstract}
We investigate the star-forming properties of \sampsize\ X-ray selected AGN host galaxies as a function of their specific X-ray luminosity (i.e., X-ray luminosity per unit host stellar mass) -- a proxy of the Eddington ratio. Our motivation is to determine whether there is any evidence of a suppression of star-formation at high Eddington ratios, which may hint toward ``AGN feedback'' effects. Star-formation rates (SFRs) are derived from fits to {\it Herschel}-measured far-infrared spectral energy distributions, taking into account any contamination from the AGN. {\it Herschel}-undetected AGNs are included via stacking analyses to provide average SFRs in bins of redshift and specific X-ray luminosity (spanning $0.01 \lesssim L_{\rm X}/M_{\ast} \lesssim 100~L_{\odot} ~M_{\odot}^{-1}$). After normalising for the effects of mass and redshift arising from the evolving galaxy main sequence, we find that the SFRs of high specific luminosity AGNs are slightly enhanced compared to their lower specific luminosity counterparts. This suggests that the SFR distribution of AGN hosts changes with specific X-ray luminosity, a result reinforced by our finding of a significantly higher fraction of starbursting hosts among high specific luminosity AGNs compared to that of the general star-forming galaxy population (i.e., 8-10 per cent vs. 3 per cent). Contrary to our original motivation, our findings suggest that high specific luminosity AGNs are more likely to reside in galaxies with enhanced levels of star-formation.
\end{abstract}



\section{Introduction}
\label{sec:introduction}

It is now widely accepted that most (if not all) massive galaxies host a central super-massive black hole (hereafter, SMBH), the masses of which display a tight correlation with the stellar masses of their host galaxy bulges \citep[e.g.,][]{Kormendy1995, Magorrian1998, Marconi2003, Ferrarese2000, Gebhardt2000, Merritt2000,Tremaine2002, Kormendy2011}. This, and the similar redshift evolution of the total star formation and SMBH accretion rate densities \citep[e.g.,][]{Silverman2008a, Aird2010, Assef2011}, are seen as two key pieces of empirical evidence of a fundamental link between SMBH and galaxy growth. However, the physical mechanisms that connect these processes remain poorly understood.

An often suggested means of connecting SMBH growth (observed as Active Galactic Nuclei, or AGN) to star formation is for the energy released by an AGN to directly influence the star formation rate (hereafter, SFR) of the host galaxy \citep[see the review of][for details on the feedback mechanism]{Fabian2012}. This could either be achieved by the AGN leading to the quenching or the enhancement of star formation. Although there is some evidence of both these scenarios in low size samples \citep[e.g.,][]{Graham1998, Klamer2004, Croft2006, Feain2007, Trichas2009, Elbaz2009, Canodiaz2012, Cicone2014}, as yet there is no consensus on how applicable these studies are to the general AGN population.

In a bid to address this, many studies have attempted to characterise the properties of AGN host galaxies and compare them to the non-AGN population. Studies based on simulations often suggest AGN act to quench star-formation \citep[e.g.,][]{DiMatteo2005, Springel2005, Hopkins2008, Zubovas2012, Dubois2013}. By contrast, most observational studies report that AGN host galaxies form a diverse population \citep[e.g.,][]{Nandra2007, Bohm2007, Schawinski2007, Silverman2008b, Xue2010, Cisternas2011, Kocevski2012}, with a preference for more massive galaxies \citep[i.e., $M_\ast$ $\gtrsim10^{10}~{M_\odot}$; e.g.,][]{Dunlop2003, Kauffmann2003, Best2005, Xue2010, Aird2012} with somewhat {\it enhanced} SFRs compared to the general population (e.g., \citealt{Barvainis2002, Priddey2003, Page2004, Schweitzer2006, Lutz2008, Hickox2009, Silverman2009, Shao2010, Bonfield2011, Santini2012, Mullaney2012a, Rosario2012, Rovilos2012, Harrison2012, Zinn2013}, although see \citealt{Page2012}). However, most studies that have explored the connection between AGN luminosity (a proxy for SMBH growth rate) and SFR have found no clear evidence of a strong correlation between these properties, at least for moderate luminosity AGNs which form the majority of the population \citep[i.e., $10^{42}<L_{X}<10^{45}~{\rm erg~s}^{-1}$; e.g.,][]{Lutz2010, Shao2010, Mullaney2012a, Rosario2012, Rovilos2012, Harrison2012, Stanley2015}. However, it is reasonable to consider that a powerful AGN (i.e., $L_{X} \sim 10^{44\--45}~{\rm erg~s}^{-1}$) will have a bigger impact on a low mass compared to a high mass galaxy, simply due to the deeper potential well of the latter. Thus, it may be more pertinent to explore the relationship between the AGN luminosity {\it relative to the host galaxy stellar mass} (hereafter, specific AGN luminosity) and the star-forming properties of the host galaxy. 

Using rest-frame ultraviolet (UV) and optical observations to measure the star-forming properties of X-ray AGN hosts, \cite{Azadi2015} recently reported that there is no evidence of a strong correlation (or anti-correlation) between specific AGN luminosity and average SFR, at least up to  $z\approx1$. Instead, they reported that the {\it fraction} of star forming galaxies increases with increasing specific AGN luminosity, suggesting a change in the underlying SFR distribution. However, \cite{Azadi2015} only probed low-to-moderate redshifts (i.e., $z<$1), thus missing the peak epoch of SMBH accretion and star formation rate densities (i.e., $z\approx2$). Furthermore, because of the potential strong contamination to the optical/UV part of the galaxy spectrum from the AGN, they were unable to consider the most luminous AGNs (i.e., $L_{X}>10^{44}~{\rm erg~s}^{-1}$), which are also those most likely to have the highest specific luminosities. Furthermore, by relying on the optical-to-UV part of the spectrum, \cite{Azadi2015} had to make considerable corrections for dust attenuation. As the problems of both AGN contamination and dust attenuation are minimal at far-infrared (FIR) wavelengths, there is significant scope to build upon the \cite{Azadi2015} study by exploiting data from the ESA {\it Herschel}\footnote{Herschel is an ESA space observatory with science instruments provided by European-led Principal Investigator consortia and with important participation from NASA.} Space Telescope.
	
Motivated by this, we extend the \cite{Azadi2015} work to more extreme redshifts (i.e., $z\approx3$) and specific X-ray luminosities (i.e., $L_{X}/M_{\ast} \sim 10 ~L_\odot~ M_\odot^{-1}$). We use a sample of \sampsize\ X-ray selected AGNs in the COSMOS, GOODS-North and GOODS-South fields, described in \S\ref{sec:dat}. In \S\ref{sec:dataprop} we describe the data analysis performed on these datasets and present the results of these analyses in \S\ref{sec:result}. We follow this with a discussion of the possible implications of our results in \S\ref{sec:discuss} before highlighting our main conclusions in \S\ref{sec:Conclusion}.

Throughout, we assume a WMAP-7 year cosmology \citep{Larson2011} and a \cite{Chabrier2003} initial mass function (hereafter, IMF) when calculating galaxy stellar masses and SFRs.


\section{Datasets}
\label{sec:dat}

Our datasets are extracted from multi-wavelength observations of three extragalactic fields: GOODS-North (hereafter, GN) and GOODS-South (hereafter, GS), and the wider but shallower COSMOS fields. This combination allows us to probe modest AGN luminosities (i.e., $L_X\sim10^{42\--43}~{\rm erg~s^{-1}}$; the dominant AGN population in terms of numbers) to $z\approx3$, while also covering the wide areas needed to include AGNs with more extreme specific luminosities (i.e., $L_{X}/M_{\ast} \sim 100 ~L_\odot~ M_\odot^{-1}$). A further benefit of using these fields are that there are the most intensely surveyed regions of the sky, with comprehensive multi-wavelength coverage spanning X-ray-to-radio regimes. In this section, we describe the datasets that we use to investigate the relationship between specific AGN luminosity and the star-forming properties of their hosts.

\subsection{X-ray data}
\label{subsec:Chandra}

Throughout, we use X-ray luminosities derived from the {\it Chandra} observations of our three fields as a proxy for bolometric AGN luminosity (using the average bolometric conversion factor from \citealt{Vasudevan2007}; see \S\ref{subsec:eddratlx}). For COSMOS, we use the 1.8 Mega seconds (hereafter, Ms) Chandra COSMOS (hereafter, C-COSMOS) survey of \cite{Civano2012} that covers the central 0.9~deg$^2$ of the COSMOS field (1761 X-ray detections in the 0.5--10 keV band). For GN, we use the 2Ms point-sources catalogue of \cite{Alexander2003} from the Chandra Deep Field North (hereafter, CDFN) survey (503 X-ray detections in the 0.5--8 keV band), while for GS we use the main 4Ms catalogue of \cite{Xue2011} from the Chandra Deep Field South (hereafter, CDFS) survey (740 X-ray detections in the 0.5--10 keV band). Merging all these fields gives a total of 3004 X-ray detected sources, which we refer to as our ``X-ray detected'' catalogue.

\subsection{Far-infrared data}
\label{subsec:fir_data}

We use the FIR emission of the host galaxy as a tracer of its SFR \citep[][]{Kennicutt1998}. Our FIR data are from instruments on board the {\it Herschel} space observatory \citep{Pilbratt2010}. The COSMOS field has been covered by {\it Herschel}'s PACS instrument \citep{Poglitsch2010} at 100~\mum\ and 160~\mum\ as part of the PACS Evolutionary Probe (PEP, \citealt{Lutz2011}) programme. For GN and GS we use the {\it Herschel}-PACS coverage resulting from combining data from the PEP and the GOODS-{\it Herschel} \citep{Elbaz2011} programmes which provide the deepest surveys of these fields at wavelengths of 100~\mum\ and 160~\mum\ \citep{Magnelli2013}. For all three fields we use the de-blended catalogues based on the positions of {\it Spitzer}-MIPS 24~\mum\ priors, which provide 100~\mum\ and 160~\mum\ fluxes for all 24~\mum\ sources, irrespective of whether they are formally detected at the longer wavelengths. We complement the PACS data with that from deep SPIRE \citep{Griffin2010} observations (at 250~\mum, 350~\mum\ and 500~\mum) of our fields carried out by (1) the HerMES programme for COSMOS and GS \citep{Roseboom2010, Oliver2012}, and (2) the GOODS-{\it Herschel} programme for GN \citep{Elbaz2011}.

For the COSMOS field, the PEP program covers all of our 1761 X-ray sources. As GOODS-{\it Herschel} did not cover the entire CDFN and CDFS we only consider X-ray sources with PACS coverage that is over 30 per cent of the maximum of each field. This results in 365 of 503 X-ray detected sources in CDFN, and 410 of 740 in CDFS, with {\it Herschel} coverage. Thus, of the 3004 X-ray sources in the three fields combined, 2536 are covered by {\it Herschel}. These form our ``{\it Herschel}-covered'' sample.

To obtain infrared counterparts for each source in our {\it Herschel}-covered sample we performed positional matching using search radii that were chosen to maximise the number of true matches, while minimising the number of spurious ones. For 100~\mum\ and 160~\mum\ fluxes we cross-matched against the {\it Herschel} catalogue using a two arc-second search radius on the 24~\mum\ prior positions. For SPIRE data in COSMOS, we cross-matched against the {\it Herschel} catalogue based on the 250~\mum\ positions, using a search radius of seven arc-seconds. For GN, SPIRE fluxes are provided in the table of \cite{Elbaz2011}, so are automatically included when we match to the PACS data. For GS we cross-matched against the {\it Herschel} catalogue using a search radius of two arc-seconds on the 24~\mum\ prior positions. Out of the 2536 X-ray sources in our {\it Herschel}-covered sample, 1178 (i.e., $\sim 46$ per cent) sources are detected in at least one {\it Herschel} band, which we refer as our ``X-ray FIR'' catalogue.

\subsection{Ancillary data}
\label{subsec:ancilliary}

As the primary goal of this study is to measure how the star-forming properties of galaxies relate to the {\it specific} X-ray luminosity (i.e., luminosity per unit host stellar mass), it is crucial that we have accurate host stellar masses for the galaxies in our sample. The most commonly used means to derive galaxy stellar masses is through fits to their broad-band spectral energy distributions (hereafter, SEDs). However, since in our case the SED can be heavily contaminated by the light from the AGN, we used a multi-component SED fitting code (i.e., {\sc CIGALE}, see \S\ref{subsec:Mstar}) to account for this contamination. As demonstrated in \cite{Ciesla2015}, {\sc CIGALE} requires UV to mid-IR photometry in order to reliably extract the stellar mass of AGN host galaxies.

Thus, we extend our datasets to include data from the UV, the optical and the near-infrared (NIR) part of the spectrum. For the COSMOS field, the UV to NIR data (i.e., {\it $u^{*}$, $B_{\rm J}$, $V_{\rm J}$, ACS606, {\it $r^+$},{\it $ i^+$}, ACS814W, {\it $z+$}, UVISTA-Y,  WFC3-F125W, UVISTA-J, WFC3-F140W, UVISTA-H, UVISTA-K$_{\rm S}$, IRAC1, IRAC2, IRAC3} and {\it IRAC4} bands) are from a combination of the \cite{Ilbert2013} catalogue and, for the central region covered by the CANDELS survey, the 3dHST catalogue compilation of \cite{Skelton2014}. For CDFN and CDFS, we use the 3dHST catalogue of \cite{Skelton2014}, that also covers the UV to the NIR wavelengths (i.e., {\it U, B, V, ACS606, R, {\it i}, {\it z}, WFC3-F125W, J, WFC3-F140W, H, K$_{\rm S}$, IRAC1, IRAC2, IRAC3} and {\it  IRAC4} bands). We cross-match these catalogues with our {\it Herschel}-covered sample using a radius of two arc-seconds on the 24 \mum\ positions when available and the X-ray positions otherwise. Combining data from all three fields, all of the 2536 sources in our {\it Herschel}-covered sample are detected in at least one of the above bands.


\section{Data analysis}
\label{sec:dataprop}

Our main focus is to investigate the star-forming properties of AGN host galaxies in terms of their specific AGN luminosities. Therefore, it is crucial that we disentangle the AGN emission from that of the host galaxy. Throughout this section, we describe how we use the X-ray fluxes to derive specific AGN luminosities, and how we account for AGN contamination when measuring host galaxy properties (i.e., stellar masses and SFRs). Furthermore, since not all of the AGNs in our {\it Herschel}-covered sample are detected in the FIR (see \S\ref{subsec:fir_data}), we also describe the stacking analysis we perform to account for these {\it Herschel} undetected sources.

\subsection{AGN properties}
\label{subsec:xrayprop}

\subsubsection{Intrinsic X-ray luminosities}
\label{subsec:xrayint}

One of the most direct ways to quantify the power of an AGN is to derive its absorption-corrected (i.e., intrinsic) X-ray luminosity. For COSMOS and GS we obtained 2--10~keV intrinsic X-ray luminosities for our {\it Herschel}-covered sample by cross-matching it with the catalogue of \citeauthor{Brightman2014} (\citeyear{Brightman2014}; hereafter, B14) which provides this information for respectively 100 per cent and 70 per cent of our sources in these two fields. To ensure consistency throughout, we also adopted the redshifts provided in B14 for these matched sources. For those sources that are not covered by B14, we derived an analytic solution to convert observed X-ray fluxes to intrinsic X-ray luminosities. We use a second-order polynomial to relate the ratio of unobscured to obscured fluxes (i.e., the flux that would be observed if there were no obscuring material relative to the actual observed flux, or ${F_{\rm unobs}/F_{\rm obs}}$) to the band ratios (i.e., Hard X-ray band/Soft X-ray band) and the redshifts ($z$) of X-ray sources in B14:

\begin{multline}
  \rm log_{10}\left(\frac{F_{unobs}}{F_{obs}}\right) = \rm \sum_{i=0}^{2} \alpha_{i} \times log_{10}\left(\frac{Hard~band}{Soft~band}\right)^{i} + \sum_{i=0}^{2} \beta_{i} \times z^{i}
  \label{eq:relation}
\end{multline}

\noindent where $\alpha_{\rm i}$ and $\beta_{\rm i}$ are polynomial coefficients. The best fit is given for $\alpha_{0}=0.23$, $\alpha_{1}=0.61$, $\alpha_{2}=0.041$, $\beta_{0}=0.01$, $\beta_{1}=-0.11$, and $\beta_{2}=-0.02$. We find that the same coefficients reproduce the intrinsic luminosities reported in B14 in both GS and COSMOS with a median of $L_{\rm Poly}/L_{\rm B14}=1.05$ and a standard deviation of 0.56.

Using Eq.\,\ref{eq:relation} together with the redshifts from \cite{Bauer2004} for the sources in GN, and the redshifts from \cite{Hsu2014} for the 30 per cent of the GS sources that are not covered by B14, we derive intrinsic 2--10~keV luminosities for all of the 2536 sources in our {\it Herschel}-covered sample.

To minimise contamination from non-AGN we only retain sources with intrinsic X-ray luminosity greater than ${\rm 10^{42}~erg\ s^{-1}}$. This has no impact on our results as we are primarily interested in high specific luminosity AGNs. Thus, we expect little contamination (i.e., typically less than 10 per cent; \citealt{Ranalli2003}) from the host galaxy to the measured X-ray flux. Among our 2536 {\it Herschel}-covered sample, 1980 are selected as true AGNs, and are referred as our ``AGN sample''.

\begin{figure}
  \centering
  \includegraphics[height=8cm, width = 9cm]{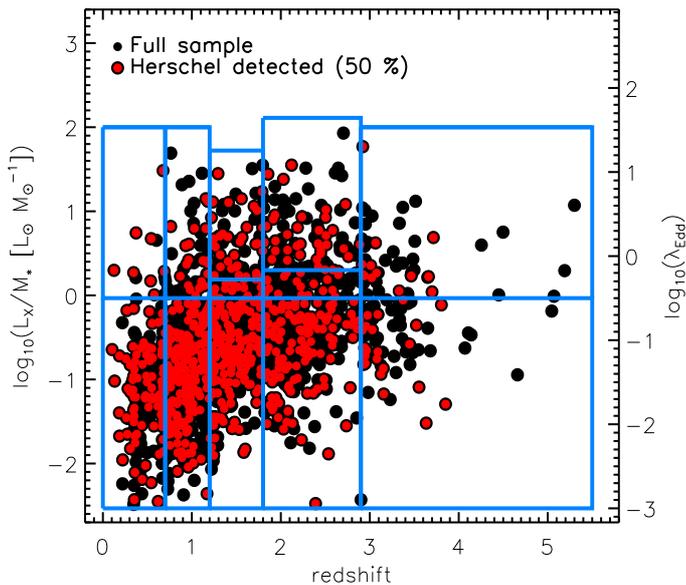}
  \caption{Specific X-ray luminosity (left-hand axis) and the corresponding Eddington ratio (righ-hand axis) versus the redshift for our final X-ray selected AGN sample from the COSMOS, CDFN and CDFS fields. The black circles represent all AGNs in our sample (i.e., X-ray detected, with stellar masses and redshift measurements), while the red circles highlight those that are also {\it Herschel} detected (i.e., in at least one {\it Herschel} band). The blue boxes show the outlines of the redshift and \protect \Ledd\ bins we use throughout our analyses. Here \Ledd\ is calculated from the specific AGN luminosity using single, fixed bolometric correction and stellar-to-SMBH mass conversion factors (see Eq.~\ref{eq:eddrat} ); hence the significant number of AGNs with log(\Ledd )$>0$ in our sample.}
  \label{fig:eddrat_z}
\end{figure}

\subsubsection{Specific L$_{\rm X}$ as a tracer of Eddington ratio}
\label{subsec:eddratlx}

The primary goal of this study is to measure the star-forming properties of AGN hosts as a function of specific AGN luminosity. For convenience, we express this specific AGN luminosity in terms of the more familiar Eddington ratio of the AGN (hereafter, \Ledd), i.e., the ratio of AGN bolometric luminosity (hereafter, $L_{\rm AGN}$) to the Eddington luminosity of its SMBH (hereafter, $L_{\rm Edd}$). To convert intrinsic X-ray luminosity to bolometric AGN luminosity we use a bolometric correction factor of 22.4 which is the median value found in \citeauthor{Vasudevan2007} (\citeyear{Vasudevan2007}; based on a local AGN sample with $L_{X} = 10^{41\--46}~{\rm erg~s^{-1}}$).  As \Ledd\ is to be used as merely a guide, we prefer to use a single bolometric correction factor over more complicated luminosity-dependent corrections \citep[e.g.,][]{Hopkins2007b}; as all our analyses are performed on the basis of specific X-ray luminosities rather than \Ledd\ the choice of bolometric correction factor has no effect on our results. For the purposes of this study, we then assume SMBH mass is proportional to the total stellar mass of the host ($M_{\ast}$, see \S\ref{subsec:Mstar} for the stellar masses), with a constant of proportionality of 0.002 \citep{Marconi2003}, giving:

\begin{equation}
\lambda_{\rm Edd}= \frac{L_{\rm AGN}}{L_{\rm Edd}} = \frac{22.4~L_{X}}{1.26 \times 10^{38} ~{\rm erg~s}^{-1}\times 0.002 \frac{M_{\ast}}{M_{\odot}}}
  \label{eq:eddrat}
\end{equation}

The distribution of our \Ledd\ sample with the redshift is shown in Fig.~\ref{fig:eddrat_z}, with the sources that are {\it Herschel} detected (i.e., at least in one of the {\it Herschel} band) shown with red points. We reiterate that we make this conversion purely to describe specific X-ray luminosities in terms of the more familiar Eddington ratio, and should not be regarded as a true measure of the Eddington ratio of the SMBH (which could explain the large numbers of AGNs with log(\Ledd) $>0$ in Fig.~\ref{fig:eddrat_z}).

\subsection{Host galaxy SFRs}
\label{subsec:firdetected}

We use the integrated 8--1000~\mum\ emission (i.e., $L_{\rm IR}$) of the AGN host galaxies as a measure of their SFRs \citep{Kennicutt1998} as it is largely unaffected by dust obscuration and AGN contamination \citep[e.g.,][]{Netzer2007, Hatziminaoglou2010}. However, it is known that a minority of powerful AGN can contribute significantly at these wavelengths, so we perform multi-component SED fitting to account for any AGN contamination to the IR luminosity. For this, we used {\sc DECOMPIR}\footnote{Publicly available and can be downloaded at\ :  \url{https://sites.google.com/site/decompir/}} \citep{Mullaney2011} that fits the IR spectral energy distribution (hereafter, SED) with an AGN component and one of five host galaxy components. The infrared luminosity of the host galaxy component can then be isolated from that of the AGN component.

We use a multi-step approach to accurately measure the SFR of the host galaxy. Firstly, we fit the SED with each of the five different galaxy templates separately, and including an AGN component. Then, to test whether or not this AGN component is required, we repeat the fit but now excluding the AGN template (allowing the host galaxy template to renormalise). Thus, for each observed IR SED, we have ten different SED models (i.e. five host plus AGN, hereafter, ``AGN+SF'', and five host only, hereafter, ``SF'').

To compare the ten different models, we cannot simply compare the chi squares of each fit because the five ``AGN+SF'' models have one degree of freedom more than the five ``SF'' models. Instead, we used the Akaike Information Criterion \citep[hereafter, AIC;][and equivalent to the Bayesian information criterion used in \citealt{Stanley2015}]{Akaike1974} that enables comparison of nested models, taking into account the number of degrees of freedom, by penalizing models with more free parameters \citep[see][for a complete analysis of the AIC]{Burnham2002}. However, because we have a low number of data points, we must use the corrected AIC (hereafter, AICc), written as:

\begin{equation}
  {\rm AICc} = \chi^{2}+2~k\left(\frac{n}{n-k-1}\right)
\end{equation}

\noindent with $\chi^{2}$ the chi square of the fit, n the number of data points, and k the number of parameters used in the model (i.e., $k_{\rm AGN+SF}$=2 and $k_{\rm SF}$=1 if we consider two and one fitting components respectively). One can see that minimizing the AICc (hereafter, AICc$_{\rm min}$) for constant {\it k} is exactly equivalent to minimizing $\chi^2$. However, rather than selecting one ``best'' model (indeed, for a finite set of templates there is no true model), we prefer to use a weighted average of all ten models, with the weighting given as:

\begin{equation}
  \mathscr{A}_{\rm w} ^{i} =\frac{\rm e^{\Delta_{AIC}^{\it i}/2}}{\sum_1^{m}{\rm e^{\Delta_{AIC}^{\it i}/2}}}
\end{equation}

\noindent with $\mathscr{A}_{\rm w} ^{i}$ the Akaike weight of the model {\it i}, $\Delta_{\rm AIC}^{i}$ the difference between the AICc of the model {\it i} and the AICc$_{\rm min}$, and {\it m} the total number of models (i.e., here {\it m}=10). 

We follow the above process for each observed SED, which gives us a weighted average {\it L}$_{\rm IR}^{\rm host}$ for each galaxy, written as:

\begin{equation}
  L_{\rm IR}^{\rm host} = \sum_{i=1}^{10}(L_{\rm {IR}~{\it i}}^{\rm host}\times \mathscr{A}_{\rm w} ^{i})
\end{equation}

$L_{\rm IR}^{\rm host}$ was then converted to SFR using eq.~4 of \cite{Kennicutt1998}, adjusted to a \cite{Chabrier2003} IMF.

\begin{table*}
  \centering
  \caption{Modules and parameter ranges used with {\sc CIGALE} to derive the stellar masses of our AGN host galaxies.}
  \def\arraystretch{1.}
  \begin{tabular}{c c}
    \hline
    \hline
    Parameter & Value\\
    \hline
    \multicolumn{2}{c|}{{\sc CIGALE} modules} \\
    \hline
    Star formation history & Delayed\\
    Stellar population synthesis model & \cite{Bruzual2003}\\
    Initial mass function & \cite{Chabrier2003}\\
    Attenuation law & \cite{Calzetti2001}\\
    Dust emission model & \cite{Dale2002}\\
    AGN & \cite{Fritz2006}\\
    \hline
    \multicolumn{2}{c|}{Range of parameters} \\
    \hline
    \multicolumn{2}{c|}{\underline{Stellar population synthesis}}\\
    e-folding time of the main stellar population (Myr) & 100, 1000, 3000, 10000, 10$^{10}$ \\
    Age of the oldest stars in the galaxy (Gyr) & 0.1, 1, 2, 3, 4, 5, 6, 7, 8, 9, 10, 11\\
    Metallicity & 0.02\\
    Age separation between the young and the old populations (Myr) & 10\\
    \multicolumn{2}{c|}{\underline{Dust attenuation}}\\
    E(B-V)* for the young population & 0.01, 0.05, 0.1, 0.2, 0.3, 0.4, 0.5, 0.6, 0.7, 0.8, 0.9, 1., 1.1, 1.2, 1.4\\
    E(B-V)* reduction factor between the old and the young populations & 0.44\\
    Central wavelength of the UV bump (nm) & 217.5\\
    Width of the UV bump (nm) & 35\\
    \multicolumn{2}{c|}{\underline{AGN}}\\
    Ratio of dust torus radii & 60\\
    Optical depth at 9.7 microns & 1.0, 6.0\\
    Beta & -0.5\\
    Gamma & 0\\
    Full opening angle of the dust torus (degrees) & 100\\
    Angle between equatorial axis and line of sight (degrees) & 0, 90\\
    Fraction of $L_{\rm IR}$ due to the AGN &  0, 0.05, 0.1, 0.2, 0.3, 0.4, 0.5, 0.7, 0.9\\
    \hline
  \end{tabular}
  \label{tab:cigale}
\end{table*}

\begin{figure}
  \centering
  \includegraphics[height=8cm, width = 9cm]{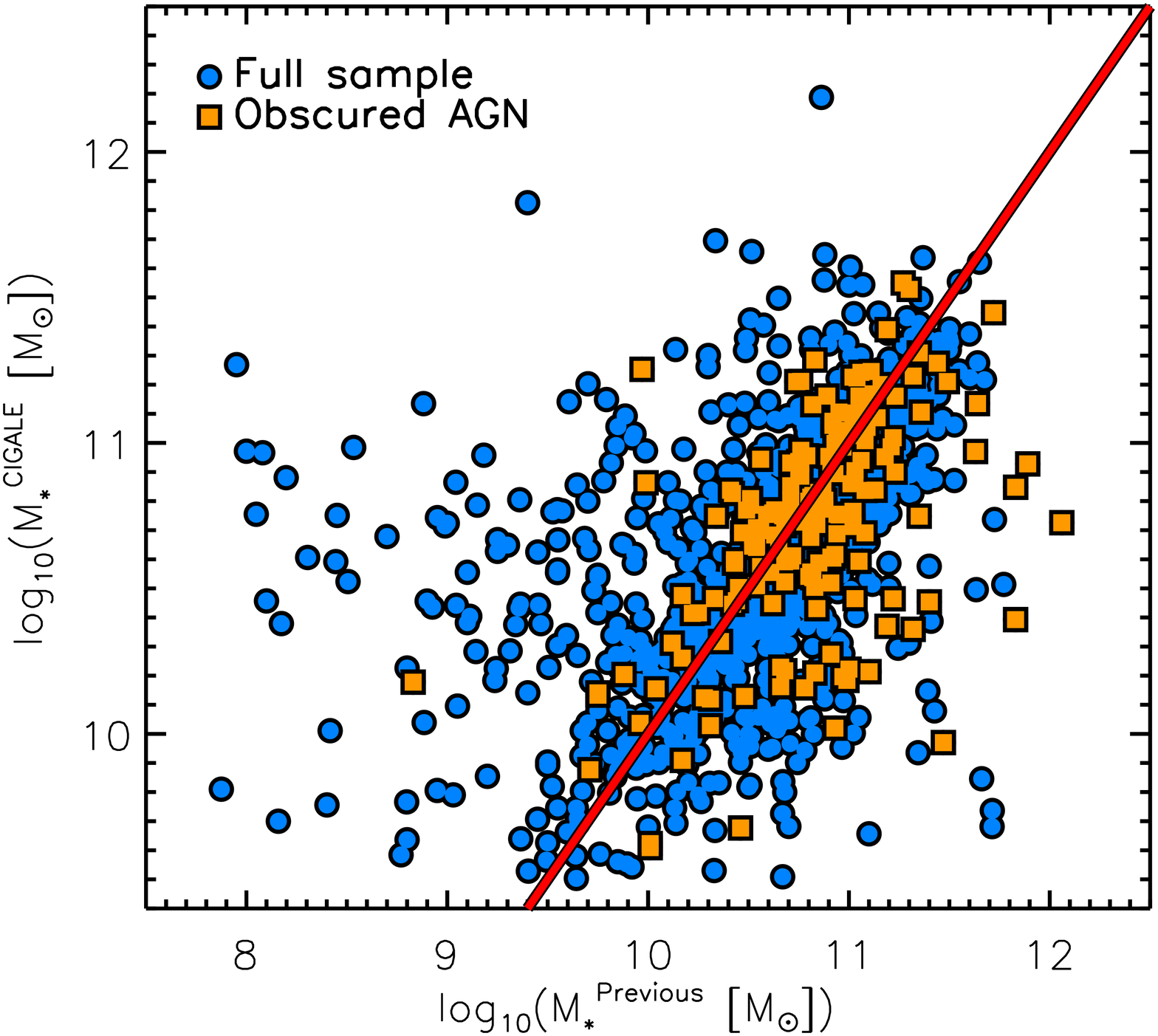}
  \caption{Comparison between the stellar masses of our AGN hosts derived using {\sc CIGALE} (i.e., $M_{\ast}^{\rm CIGALE}$) and the stellar masses given in \protect \cite{Ilbert2013} and \protect \cite{Elbaz2011} tables for COSMOS and GOODS, respectively (i.e., $M_{\ast}^{\rm Previous}$). The latter do not account for an AGN component when fitting the SEDs. The blue circles represent all AGNs of our sample (i.e., X-ray detected, with stellar masses and redshift measurements) with a matching $M_{\ast}^{\rm Previous}$. The orange squares represent optically obscured AGNs. The one-to-one line is shown in red.}
  \label{fig:Mstar_comp}
\end{figure}

\subsection{Stellar masses of the AGN host galaxies}
\label{subsec:Mstar} 

Our main goal is to study how star-forming properties of AGN hosts change as a function of specific AGN luminosity. As we define specific luminosity as luminosity {\it per unit host stellar mass}, it is crucial that we accurately derive host stellar masses. In the presence of an AGN the optical to NIR wavelengths -- from which stellar masses are normally derived -- can suffer contamination by light from the AGN \citep[e.g.,][]{Stern2005, Donley2012}. To account for this contamination we used the multi-component SED fitting code {\sc CIGALE} \footnote{{\sc CIGALE} is publicly available at \url{http://cigale.lam.fr/}} that includes an AGN component to the fit \citep[][Boquien in preparation, Burgarella in preparation]{Ciesla2015}. Recently, \cite{Ciesla2015} demonstrated that {\sc CIGALE} is able to reproduce the stellar masses of mock galaxies (generated using the {\sc GALFORM} code; \citealt{Cole2000}) to within 40 per cent for galaxies where up to 70 per cent of the optical to NIR emission is due to the AGN (only $\sim 1$~per cent of our AGNs are above an AGN contamination of 70 per cent at those wavelengths).

The parameters that control {\sc CIGALE}'s fitting process were selected based on what \cite{Ciesla2015} found best reproduced the stellar masses of their mock galaxy samples (see Table~\ref{tab:cigale}). The uncertainty on stellar mass increases with decreasing host stellar mass due to a combination of faintness and higher relative AGN contributions. Therefore, we only select AGNs with host galaxy stellar masses satisfying $M_{\ast}>10^{9.5}~M_{\odot}$. Within our AGN sample of 1980 sources, \sampsize\ satisfy this criterion. The level of AGN contamination spans the range 5--85~per cent. However, 95~per cent of the sample have AGN contamination of less than 50 per cent \citep[within which][report that the uncertainty on the stellar mass is less than 20~per cent]{Ciesla2015}. We also note that the 5~per cent of AGNs in our sample with AGN contamination above 50~per cent is distributed across our \Ledd\ range and thus do not adversely effect one region of parameter space relative to another.

As a consistency check, we compared the AGN host stellar masses derived using {\sc CIGALE} ($M_{\ast}^{\rm CIGALE}$) to those reported in \cite{Ilbert2013} for COSMOS and in \cite{Elbaz2011} for GN and GS ($M_{\ast}^{\rm Previous}$). Both these other studies take steps to account for AGN contamination, but use more simplistic approaches than our SED-fitting based method. Specifically, for galaxies whose {\it Spitzer}-IRAC bands are AGN dominated (flagged by being characterised by a power-law),  \cite{Ilbert2013} ignore these bands when deriving stellar masses, whereas \cite{Elbaz2011} remove them from their sample entirely. Thus, by comparing our masses with those of \cite{Ilbert2013}, we obtain a measure of the discrepancy between different approaches of accounting for AGN contamination when calculating host stellar masses. This comparison is shown in Fig.~\ref{fig:Mstar_comp} with filled blue circles. The standard deviation of the log($M_{\ast}^{\rm {\sc CIGALE}}$/$M_{\ast}^{\rm Previous}$) ratios of the full sample is 0.5~dex and there is a systematic offset of 0.31~dex (i.e., a factor of $\sim2$). However, we also separate obscured (type 2) and unobscured (type 1) AGNs in this Figure using optical classifications for GN and GS. When we do this, we find a smaller standard deviation and systematic offset (i.e., 0.4~dex and 0.09~dex, respectively) for obscured AGNs. This suggests that the offset in the full sample arises from our different treatments of AGN contamination, which will be stronger for unobscured AGNs. Therefore, we do not correct for any offset, believing that our treatment of AGN contamination by including AGN templates in our SED fits is more robust than those used by \cite{Elbaz2011} and \cite{Ilbert2013}, i.e., studies where AGN were not the primary focus.

\begin{table*}
  \centering
  \caption{Description of our redshift and \protect \Ledd\ bins.}
  \def\arraystretch{1.5}
  \begin{tabular}{c|*{2}{c|} c c c c c }
    \hline
    \hline
    (1) & \multicolumn{2}{c|}{(2)} & (3) & (4) & (5) & (6) & (7) \\
    Index & \multicolumn{2}{c|}{Bin description} & N & $z$ & $M_{*}$ ($\times 10^{10}$~$M_{\odot}$) & \Ledd\ & $L_{X}/M_{\ast}~(L_{\odot}~M_{\ast}^{-1})$\\
    \hline
	&      \multicolumn{2}{c|}{All}   & \sampsize\ & 1.57 & 4.16 & 0.10 & 0.70\\
    \hline
    1 & z = 0--0.7;   &	log(\Ledd\ )$<$-0.5 & 170&0.50 &4.68$\pm$0.05 & 0.05$\pm$0.01  & 0.13$\pm$0.01  \\
    2 & z = 0--0.7;   &	log(\Ledd\ )$>$-0.5 & 11 &0.51 &2.17$\pm$0.13 & 1.64$\pm$0.01  & 4.81$\pm$0.29  \\
    \hline
    3 & z = 0.7--1.2;   &	log(\Ledd\ )$<$-0.5 & 398 &0.94 &6.22$\pm$0.04 & 0.06$\pm$0.01  & 0.18$\pm$0.01  \\
    4 & z = 0.7--1.2;   &	log(\Ledd\ )$>$-0.5 & 47 &1.00 &3.12$\pm$0.09 & 1.97$\pm$0.07  & 5.76$\pm$0.21  \\
    \hline
    5 & z = 1.2--1.8;   &	log(\Ledd\ )$<$-0.5 & 323 &1.47 &6.47$\pm$0.05 & 0.11$\pm$0.01  & 0.31$\pm$0.01  \\
    6 & z = 1.2--1.8;   & -0.5$<$log(\Ledd\ )$<$-0.28 & 47&1.48 &4.63$\pm$0.11 & 0.41$\pm$0.01  & 1.20$\pm$0.04  \\
    7 & z = 1.2--1.8;   & log(\Ledd\ )$>$-0.28 & 67 &1.51 &2.68$\pm$0.06 & 2.18$\pm$0.10  &  6.38$\pm$0.28 \\
    \hline
    8 & z = 1.8--2.9;   & log(\Ledd\ )$<$-0.5 & 285 &2.22 &9.64$\pm$0.07 & 0.13$\pm$0.01  & 0.37$\pm$0.01  \\
    9 & z = 1.8--2.9;   &  -0.5$<$log(\Ledd\ )$<$-0.17 & 74 &2.27 &4.99$\pm$0.14 & 0.47$\pm$0.01  & 1.37$\pm$0.04  \\
    10 & z = 1.8--2.9;   &  log(\Ledd\ )$>$-0.17 & 84 &2.33 &2.62$\pm$0.08 & 2.95$\pm$0.12 & 8.63$\pm$0.34  \\
    \hline
    11 & z $>$ 2.9;   &  log(\Ledd\ )$<$-0.5 & 69 &3.29 &14.3$\pm$0.17 & 0.12$\pm$0.01  & 0.33$\pm$0.01  \\
    12 & z $>$ 2.9;   &  log(\Ledd\ )$>$-0.5 & 45 &3.36 &4.66$\pm$0.13 & 1.70$\pm$0.08 & 3.96$\pm$0.22  \\
    \hline
    \multicolumn{8}{p{14.5cm}}{\footnotesize{\textit{Notes}: (1)  Bin index. (2) redshift and \protect \Ledd\ ranges probbed in the bin. (3) Number of AGNs. (4) Average redshift. (5) Average stellar mass. (6) Average Eddington ratio. (7) Corresponding average specific X-ray luminosity.}}
  \end{tabular}
  \label{tab:stack}
\end{table*}

\begin{figure*}
  \centering
  \includegraphics[scale = 0.3]{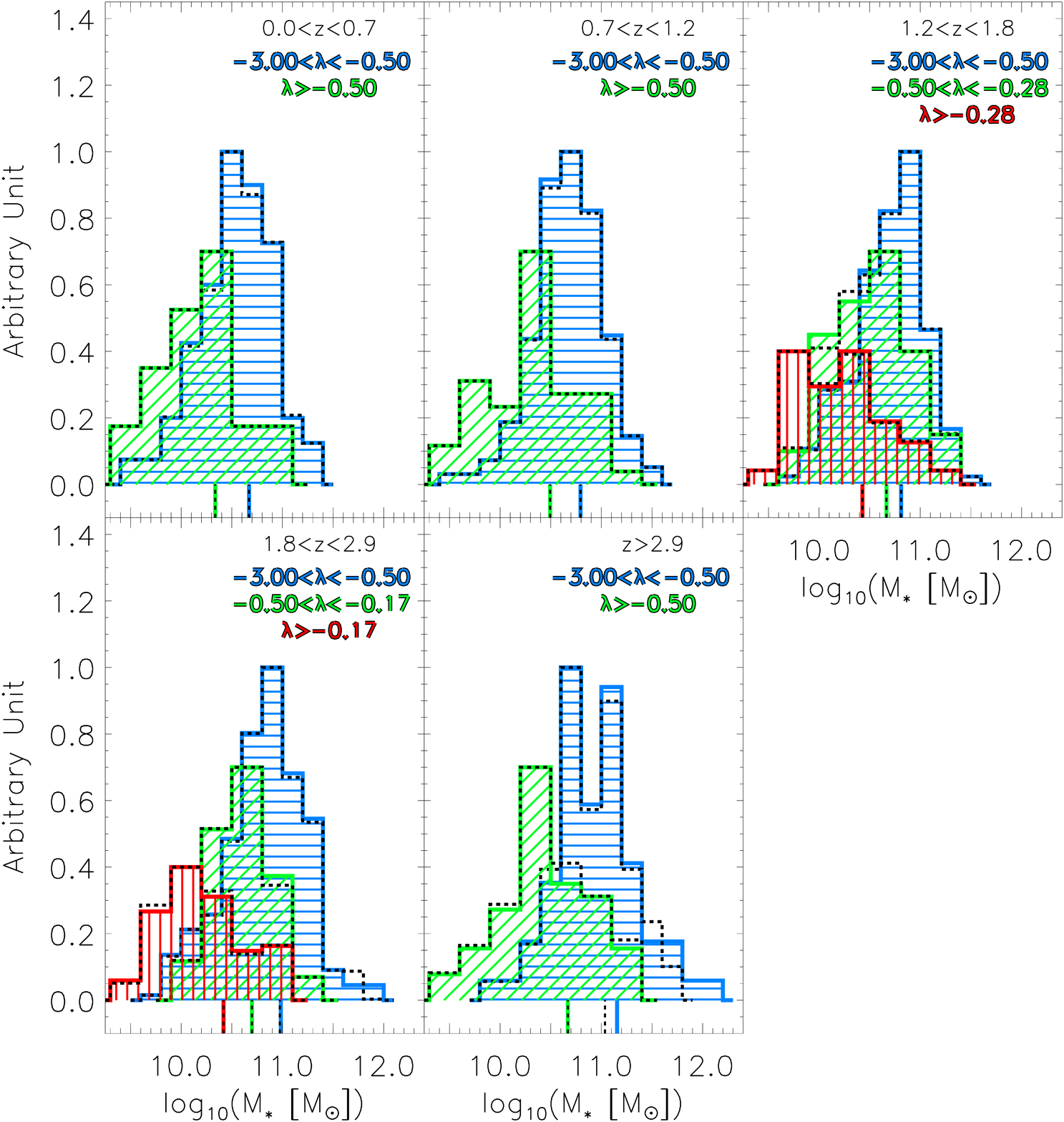}
  \caption{The distribution of AGN host stellar masses in each of our redshift and \protect \Ledd\ bins (see key), shown as coloured histograms. The black dashed-lines show the corresponding stellar-mass distributions of our mass-matched sample (which often lie over the coloured histograms). The average AGN host stellar-masses are shown as vertical lines underneath each distribution (adopting the same colour-coding as the histograms). Similarly, the corresponding average stellar-masses in the mass matched sample are shown with dashed-lines underneath each distribution (again, which often lie over the coloured vertical lines).}
  \label{fig:massdistrib}
\end{figure*}

\subsection{FIR stacking analysis}
\label{subsec:stacking} 

Despite using the deepest FIR data available for our fields, only around $\sim$20 per cent of our AGN sample is detected in enough {\it Herschel} bands to perform SED fitting (i.e., in at least three bands; see \S\ref{subsec:fir_data}). To incorporate the remaining 80 per cent of sources, we perform stacking analysis to derive their average infrared SEDs. This involves co-adding detected and undetected sources at the positions of the X-ray AGNs and measuring the total resulting flux. By dividing this total by the number of AGNs in the stack, we obtain the average flux of the stacked AGNs. Here, we separately stack the six (far-)infrared bands: 24~\mum, 100~\mum, 160~\mum, 250~\mum, 350~\mum\ and 500~\mum, using the 24~\mum\ positions when detected in this band (i.e., 72 per cent), and the X-ray positions otherwise.

Prior to stacking, we split our sample into bins of specific AGN luminosity and redshift (see Table~\ref{tab:stack}). This is to enable us to determine how the average star-forming properties of the host galaxies change with these parameters. To minimise shot noise we aim to have over 400 AGNs in each of our redshift bins. However, this is impossible for our lowest and highest redshift bins, which still contain over 100 AGNs. Within each of our redshift bins, we also bin according to Eddington ratio (or, in reality, specific X-ray luminosity). In each redshift bin, all AGNs with -3.0$<$log(\Ledd )$<$-0.5 are binned together to represent the low \Ledd\ regime. This is because this low \Ledd\ regime has already been explored in \cite{Azadi2015}, at least at $z<1$. This low \Ledd\ bin includes most of our X-ray selected AGNs at all redshifts (i.e., 95 per cent, 91 per cent, 78 per cent, 67 per cent and 60 per cent for $z=0\--0.7, 0.7\--1.2, 1.2\--1.8, 1.8\--2.9$, and $z>2.9$, respectively). The high Eddington ratio regime (i.e., log(\Ledd) $>$ −0.5) is then split into one or two bins, depending on the number of AGNs in each redshift bin (i.e., the two redshift bins containing over 100 AGNs with log(\Ledd)$ >$-0.5, i.e., z = 1.2 -- 1.8 and 1.8 -- 2.9, have two high Eddington ratio bins). Our various bin boundaries are shown in Fig. 1 and summarised in Table 2.

We perform stacking analysis on each of our redshift and \Ledd\ bins, following the procedure outlined in \cite{Schreiber2015}. Briefly, we cut out 51x51 pixel sub-images of the science maps centred on the positions of all the AGNs in each of our bins (i.e., whether detected or undetected in that band). We then sum these sub-images and divide the resulting image by the number of AGNs in the stack. The average flux of the AGNs in the bin is then measured by fitting the instrument point spread function (hereafter, PSF) at the centre of the stacked cutout. We extract the flux within a circular aperture and perform aperture corrections. We then measure the background flux in an annulus drawn around the source and subtract it from our measured source flux. The aperture sizes, annulus radii and aperture corrections we used for all the various bands are reported in the observer manuals of the respective instruments.\footnote{The IRAC instrument handbook is publicly available at:\\ \url{http://irsa.ipac.caltech.edu/data/SPITZER/docs/irac/}}
\footnote{The PEP full public data release handbook is publicly available at:\\ \url{http://www.mpe.mpg.de/resources/PEP/DR1_tarballs/readme_PEP_SPIRE.pdf}}
\footnote{The SPIRE instrument handbook is publicly available at:\\ \protect{\url{http://herschel.esac.esa.int/Docs/SPIRE/spire_handbook.pdf}}}
We took a bootstrapping approach to estimate the error on the stacked flux in each band by randomly re-sampling one third of the AGNs in each bin (with replacement) and performing the above stacking analyses. This was performed 100 times for each bin and for each band, with the standard deviation of these 100 trials divided by the square root of three taken as the uncertainty in the flux.

When stacking, any contributing flux from bright neighbors can boost the true flux of the stacked source. This contribution will vary from source to source. If the sources are randomly distributed across the field then the level of boosting will be roughly the same in each stack (if the number of stacked sources is large to overcome shot noise). However, in presence of clustering the probability of finding neighbors close to the stacked sources increases \citep{Chary2010}, and meaning the level of boosting increases with the level of clustering. To account for this effect we adopt the approach of \cite{Schreiber2015}.  Briefly, the effects of clustering is measured by stacking on simulated maps where all the input fluxes are known. These maps are generated by producing a mock galaxy sample with a specific SFR distribution described by \cite{Sargent2012}. The SFR of these galaxies are then calculated using the known stellar masses of the galaxies. The flux distribution of these galaxies is then derived using the known redshifts of the galaxies (i.e., K-correction). Sources with these fluxes are then placed on the noise map (generated from the root mean square map of the field) at the positions of the detected sources. The simulated maps therefore have sources with the same level of clustering as the science maps, but which have different (i.e., simulated) fluxes. We then stack sources on this simulated map in bins of redshift and stellar mass. The contamination due to clustering ($\sigma_{\rm clus}$) is then calculated as the difference between the stacked flux and the known mean flux of the simulated sources. We performed 100 of realisations (generating a new map each time), the position of the peak of the resulting $\sigma_{\rm clus}$ distribution gives the average contribution due to clustering. The results of our simulation are in agreement with \cite{Schreiber2015} (i.e.,  $\sigma_{\rm clus}$ = 0 per cent, 0 per cent, 3 per cent, 8 per cent, 13 per cent, 25 per cent in the 24~\mum , 100~\mum , 160~\mum , 250~\mum , 350~\mum\  and 500~\mum\ band, respectively). We apply these corrections to our stacked fluxes to take into account the clustering bias.

Finally, average SFRs are calculated from the stacked fluxes by using the same SED-fitting approach that individually detected sources (see \S\ref{subsec:firdetected}). The uncertainty on the average SFR is calculated by adopting a Monte Carlo approach, i.e., adding Gaussian noise to the stacked fluxes (according to the size of uncertainty on the fluxes) and re-performing the SED fits. Average SFRs and their associated errors are given in the column 2 of the Table~\ref{tab:results}.

When using FIR wavelengths to derive SFRs, we assume that the dust re-processes all the UV at the FIR wavelengths. However, a non-negligible amount of UV light could also be directly emitted (i.e., not re-processed by dust), which we estimate using the average relation between the attenuation (i.e., $L_{\rm IR }/L_{\rm UV}$) and the stellar mass reported in \cite{Heinis2014}. This is then converted into SFR$_{\rm UV}$ using the \cite{Kennicutt1998} relation (modified for a \citealt{Chabrier2003} IMF) and added to the SFR$_{\rm IR}$ to give the total SFR of the galaxy. Taking this approach, we find that the UV component represents between 1 to 10 per cent of the total SFR, depending on the average stellar mass of the bin. In what follows, all the SFRs are the total SFRs (i.e., SFR$_{\rm IR}+\rm SFR_{\rm UV}$).

Finally, using eq.~15 of \cite{Ranalli2003} to calculate the 2--10~keV flux arising from star-formation in the host galaxy, we also verified that any contamination to the X-ray luminosities from the host galaxies are negligible; in each of our bin we found a contamination which is less than 1 per cent.

\subsection{Mass matched sample}
\label{subsec:massmatch}

To explore the dependence of star formation on the specific X-ray luminosity of AGNs, and how this may have evolved with redshift, we bin our sample in terms of these parameters (see \S\ref{subsec:stacking}). The AGNs in each of our bins follow a different stellar mass distribution (see color shaded histograms in Fig.~\ref{fig:massdistrib}). With SFR related to stellar mass via the galaxy main sequence \citep[hereafter, MS; e.g.,][]{Salim2007, Daddi2007, Rodighiero2011, Elbaz2011, Sargent2012, Schreiber2015}, it is important to take into account for any differences in the stellar mass distributions when comparing between bins. To achieve this we calculate the difference between the measured average SFR of the AGNs in each bin and the average the sample would have if all the host galaxies lay on the MS. To calculate the latter, we would ideally use a large sample of mass-matched non-AGN, MS galaxies. However, because luminous AGNs (i.e., of the type explored here) tend to reside in high mass galaxies \citep[e.g.,][]{Dunlop2003, Kauffmann2003, Best2005, Aird2012}, a sufficiently large enough sample of real high mass with the same level of {\it Herschel} coverage does not exist. Instead, following \cite{Bernhard2014} we generate a population of 25,002,047 mock MS galaxies and assign SFRs to these galaxies using the evolving specific SFR distribution of MS galaxies reported in \citeauthor{Schreiber2015} (\citeyear{Schreiber2015}; this represents an update to the procedure outlined in \citealt{Bernhard2014} where the evolving MS of \citealt{Rodighiero2011} was used). As such, this sample of mock galaxies has exactly the same SFR distribution as a sample of MS galaxies, but is significantly larger than any observed sample of galaxies. Thus, for each of our AGN, we can match 5 mock galaxies with the same stellar mass and redshift (to within $\Delta_{z}=0.1$ and $\Delta_{\rm log({\it M})}=0.005$). The dotted lined histograms in Fig.~\ref{fig:massdistrib} show the corresponding mass-matched sample for each of our redshift and \Ledd\ bins. We then calculate the mean SFR of the mock sample for each of our bins, with the uncertainty on the mean calculated by randomly resampling one third of the mock population 100 times and measuring the spread of resulting distributions of means.

\begin{figure}
  \centering
  \includegraphics[scale = 0.16]{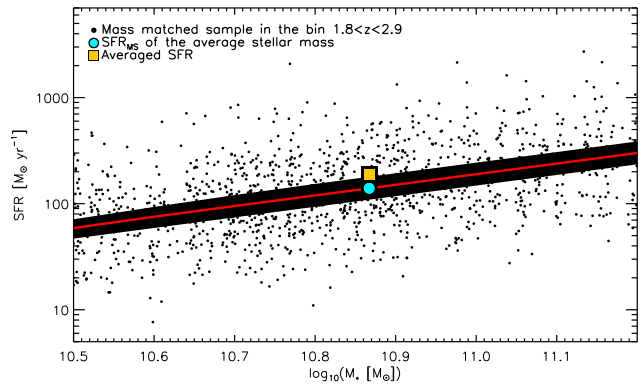}
  \caption{SFR versus stellar-mass for our mass-matched sample in the $1.8<z<2.9$ redshift bin. The black area shows the evolution of the MS from $z=1.8$ to $z = 2.9$ (from \protect \citealt{Schreiber2015}) while the red line represents the average position over this redshift range. The blue circle displays the SFR$_{\rm MS}$ corresponding to the mean stellar-mass of this bin while the orange square displays the mean SFR of the mass-matched sample. The latter represents a true indication of the average SFR of the sample, and is offset by a factor of $\approx 1.5$ above that derived by converting from the mean stellar mass of the bin.}
  \label{fig:matchdif}
\end{figure}

The benefits of using a mass-matched sample over simply assuming the average stellar mass of the observed AGNs to calculate the average MS SFR is shown in Fig.~\ref{fig:matchdif} for the bin $1.8<z<2.9$. Here, it is clear that simply taking the average stellar mass and converting that to an average MS SFR (using, for example, eq. 9 in \citealt{Schreiber2015}) results in an average SFR that is a factor of $\approx$1.5 below that calculated from using a mass-matched sample.

\begin{figure*}
  \centering
  \includegraphics[scale = .30]{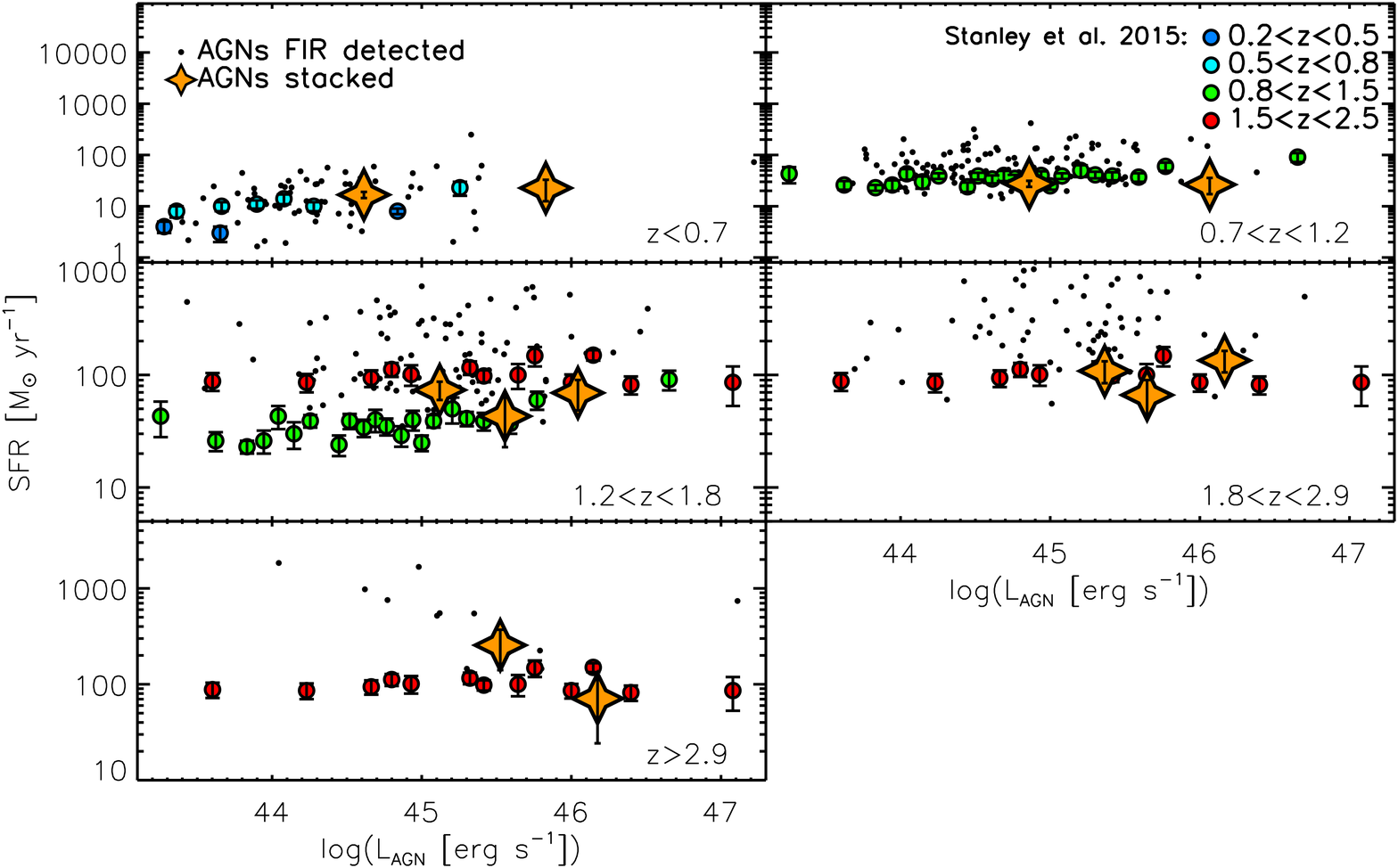}
  \caption{SFR for our X-ray selected AGN hosts versus AGN luminosity. Each panel represents a different redshift bin. The black dots represent the AGNs that are individually-detected by {\it Herschel}, whereas the orange stars show the average SFRs derived from the stack. As a comparison, we also show the results of \protect \cite{Stanley2015}, colour coded according to their redshift bins. We find that our average SFRs are broadly consistent with theirs.}
  \label{fig:sfrlx}
\end{figure*}

\begin{figure*}
  \centering
  \includegraphics[scale = .30]{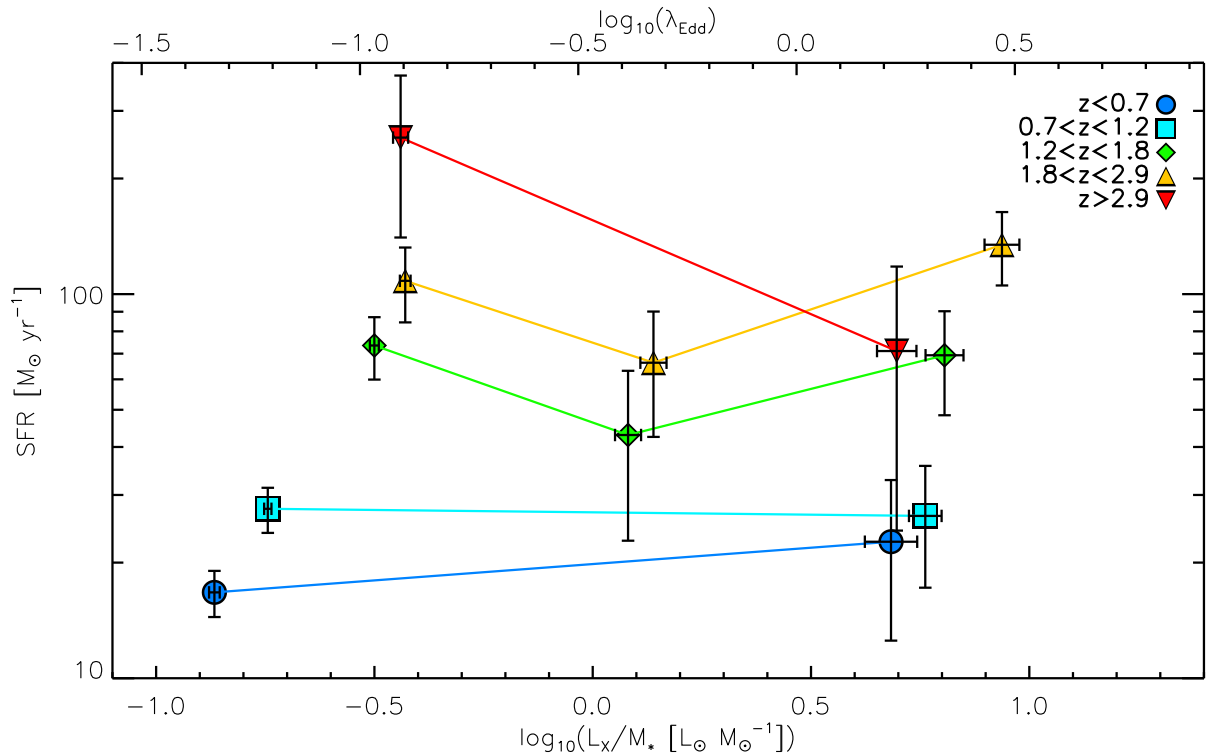}
  \caption{SFRs of our sample of AGN host galaxies versus specific X-ray luminosity (bottom axis) or \protect \Ledd\ (top axis). Coloured symbols show average host SFRs in each of our redshift and \protect \Ledd\ bins (see key). The error bars show the 1$\sigma$ error on the mean value. We find no evidence of a strong correlation between SFR and \protect \Ledd .}
  \label{fig:sfreddrat}
\end{figure*}

\begin{table*}
  \centering
  \caption{Average SFRs, normalised SFRs and minimum and maximum fractions of starbursts in each of our redshift and \protect \Ledd\ bins.}
  \def\arraystretch{1.5}
  \begin{tabular}{c c c c}
    \hline
    \hline
      (1) & (2) & (3) & (4) \\
    Index & SFR ($M_{\odot}~{\rm yr}^{-1}$) & SFR/SFR$_{\rm MS}$ & Fraction [minimum - maximum] (percent) \\
    \hline
    1 & 16.73$\pm$2.30 & 0.43$\pm$0.06 & 2 -- 3 \\
    2 & 22.67$\pm$10.2 & 1.07$\pm$0.54 & 8 --8 \\
    \hline
    3 &27.61$\pm$3.71 & 0.36$\pm$0.05 & 1 -- 6 \\
    4 & 26.46$\pm$9.25 & 0.64$\pm$0.23 & 33 -- 10\\
    \hline
    5 & 73.50$\pm$13.6 & 0.61$\pm$0.12 & 6 -- 29 \\
    6 & 42.98$\pm$20.2 & 0.58$\pm$0.28 & 6 -- 33 \\
    7 & 69.31$\pm$20.9 & 1.25$\pm$0.40 & 14 -- 66\\
    \hline
    8 & 108.3$\pm$23.9 & 0.45$\pm$0.10 & 3 -- 50\\
    9 & 66.28$\pm$23.8 & 0.52$\pm$0.19 & 7 -- 60 \\
    10 & 134.4$\pm$29.1 & 2.30$\pm$0.53 & 10 -- 84 \\
    \hline
    11 & 255.6$\pm$115.0 & 0.63$\pm$0.29 & -- \\
    12 & 71.10$\pm$46.9 & 0.35$\pm$0.23 & -- \\
    \hline
    \multicolumn{4}{p{10.5cm}}{\footnotesize{\textit{Notes}: (1) Bin index (as in Column 1 of Table~\ref{tab:stack}). (2) Average SFR measured after removing the AGN contamination. (3) Average normalised SFR accounting for the stellar mass bias.(4) Minimum and maximum fraction of AGN host that are starburst.}}
  \end{tabular}
  \label{tab:results}
\end{table*}

\begin{figure*}
  \centering
  \includegraphics[scale = .30]{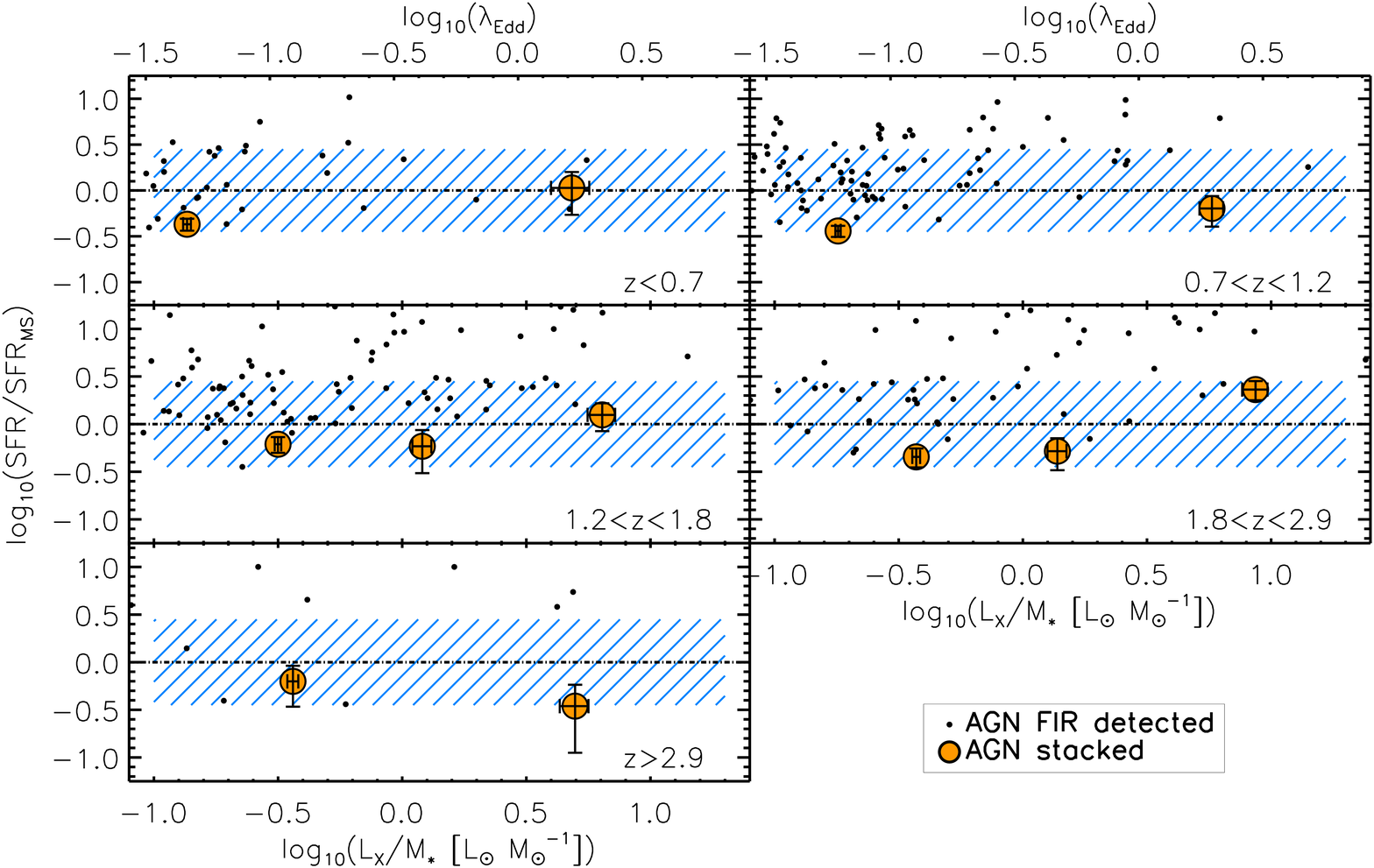}
  \caption{Mass and redshift-normalised SFRs of our sample of AGN host galaxies versus specific X-ray luminosity (bottom axis) or \protect \Ledd\ (top axis). Each panel represents a different redshift bin. The black dots represent AGNs that are individually detected by {\it Herschel}, whereas the orange circles represent average normalised SFRs. The black dashed line indicates the average position of MS galaxies; its $1\sigma$ scatter is shown as the blue shaded area. All the error bars show the 1$\sigma$ error on the mean value. We find that AGNs in our highest \protect \Ledd\ bins at $z=1.8-2.89$ has higher normalised SFRs compared to their lower \Ledd\ counterparts at a significance of 3.13$\sigma$.}
  \label{fig:lagnsfr}
\end{figure*}


\section{Results}
\label{sec:result}

In the previous sections, we described how we arrived at our final sample of \sampsize\ AGNs with {\it Herschel} coverage and reliable host stellar masses, and how we measured the star-formation rates of their host galaxies. In this section, we describe how we used this dataset to explore the star-forming properties of the AGN hosts as a function of specific X-ray luminosity (equivalently, \Ledd ; see \S\ref{subsec:eddratlx}), and whether any relation between these properties has evolved with redshift. First, however, we check that our average SFRs are consistent with those reported in previous studies.

\subsection{SFR as a function of total X-ray luminosity}
\label{subsec:sfrxray} 

Most previous studies that have explored the connection between SMBH and galaxy growths have found no clear evidence for a strong correlation between the total X-ray luminosity and the SFR of the host galaxy, particularly for the dominant population of moderate luminosity AGNs \citep[i.e., $10^{42} < L_{X} < 10^{45}~{\rm erg~s^{−1}}$; e.g.,][]{Lutz2010, Shao2010, Mullaney2012a, Rosario2012, Rovilos2012, Harrison2012, Stanley2015}. To check whether our data analyses are producing reliable results, we compare our measured SFRs against those reported in one of the most recent of such studies \citep[i.e.,][]{Stanley2015}. This comparison is shown in Fig.~\ref{fig:sfrlx}, where we plot our measured SFRs alongside those from \cite{Stanley2015}. We have fewer bins than \cite{Stanley2015} as we bin in terms of \Ledd\, rather than $L_{X}$ as used in their study. However, we calculated the average AGN luminosity of each of our bins using Eq.~\ref{eq:eddrat} to enable us to perform a direct comparison. For all of our redshift bins, our measured average SFRs (i.e., the yellow stars in Fig.~\ref{fig:sfrlx}) are in agreement with \cite{Stanley2015} over the range of AGN luminosities covered by our study (i.e. $ 10^{43} < L_{\rm AGN} < 10^{45} ~{\rm erg~s^{−1}}$ ). As found in \cite{Stanley2015}, our results indicate that there is no evidence of a strong correlation between average SFR and total X-ray luminosity, at least for AGN luminosities covered in this study. Our results also reconfirm that average SFRs of AGN host galaxies increase with redshift, consistent with suggestions that average SFRs of AGN hosts are similar to those of non-AGN MS galaxies.

\subsection{Star-forming properties as a function of specific X-ray luminosity}
\label{subsec:lagnsfr} 	

Motivated by the absence of a strong correlation between the total X-ray luminosity and the SFR of the host galaxy (see \S\ref{subsec:sfrxray}), we test whether any correlation exists between the specific X-ray luminosity (i.e., the \Ledd ), and the average SFR. In Fig.~\ref{fig:sfreddrat} we show how the stacked SFR for our X-ray selected AGN sample change with specific X-ray luminosity (bottom axis in Fig.~\ref{fig:sfreddrat}), or with \Ledd\ (top axis in Fig.~\ref{fig:sfreddrat}), in each of our different redshift bins (color coded in Fig.~\ref{fig:sfreddrat}). We find that at fixed \Ledd\ the average SFR increases with the redshift. However, at fixed redshift, there is no evidence of a strong correlation between average SFR and \Ledd .

Not shown in Fig.~\ref{fig:sfreddrat}, however, is the effect of the stellar mass gradient along the abscissa. Due to the X-ray flux limit of our sample, AGNs with low \Ledd\ will tend to have higher stellar masses compared to the high \Ledd\ AGNs (see the stellar mass histograms in Fig.~\ref{fig:massdistrib}). Since SFR is related to stellar mass via the MS, this gradient may have an effect on the observed relationship between SFR and \Ledd , such that higher \Ledd\ AGNs (i.e., in galaxies with preferentially lower stellar masses) should tend to have lower SFRs (i.e. by roughly a factor of $\sim$5 according the observed stellar mass gradient). Thus, the observed flat relationship between SFR and \Ledd\ suggests that the hosts of high \Ledd\ AGNs have boosted SFRs (relative to the MS) compared to low \Ledd\ AGNs. 

In light of the above, we use our mock mass-matched sample of MS galaxies to normalise for the effects of the mass gradient within our samples. To do this we take the measured mean SFRs of each of our bins and divide them by the mean SFR of the corresponding mass-matched MS sample (i.e., ${\rm SFR_{MS}}$; see \S\ref{subsec:massmatch}). Uncertainties on this ratio were derived by propagating the errors on the mean SFR and ${\rm SFR_{MS}}$ (see \S\ref{subsec:stacking} and \S\ref{subsec:massmatch}). In Fig.~\ref{fig:lagnsfr} we display how the normalised average SFR (orange filled circles in Fig.~\ref{fig:lagnsfr}) evolves with specific X-ray luminosity (bottom axis in Fig.~\ref{fig:lagnsfr}), or \Ledd\ (top axis in Fig.~\ref{fig:lagnsfr}) at different redshifts. In this figure we also show the $3\sigma$ range of non-AGN MS galaxies as reported in \citeauthor{Schreiber2015} (\citeyear{Schreiber2015}; blue shaded area). Our results indicate that at low \Ledd\ (i.e., log(\Ledd )$<-0.5$) AGN hosts display SFRs toward the lower edge of the MS range at all redshifts. By contrast, our results suggest that AGNs with log(\Ledd\ )$>0$ show slightly enhanced average  normalised SFRs compared to log(\Ledd )$<-0.5$ AGNs, by factors of 2.25$\pm$1.18, 1.75$\pm$0.67, 2.04$\pm$0.75, 5.11$\pm$1.63 for the redshift bins $z<0.7, 0.7<z<1.2, 1.2<z<1.8~{\rm and}~ 1.8<z<2.9$, respectively. However, it must be noted that it is only in the 1.8$<z<$2.9 redshift bin that this increase is measured at a significance of $>3\sigma$ (i.e., 3.13$\sigma$).  This represents the bin containing the largest number of high Eddington ratio AGNs (i.e., 84 AGNs in the bin -0.17$<$log(\Ledd )$<$1.64), so it is possible that larger samples may increase the significance of our other findings. It should also be noted, however, that the significance of any difference drops to $<3\sigma$ in all our redshift bins if we exclude galaxies with $M_{\ast}<10^{10}M_\odot$.

\begin{figure*}
  \centering
  \includegraphics[scale = .30]{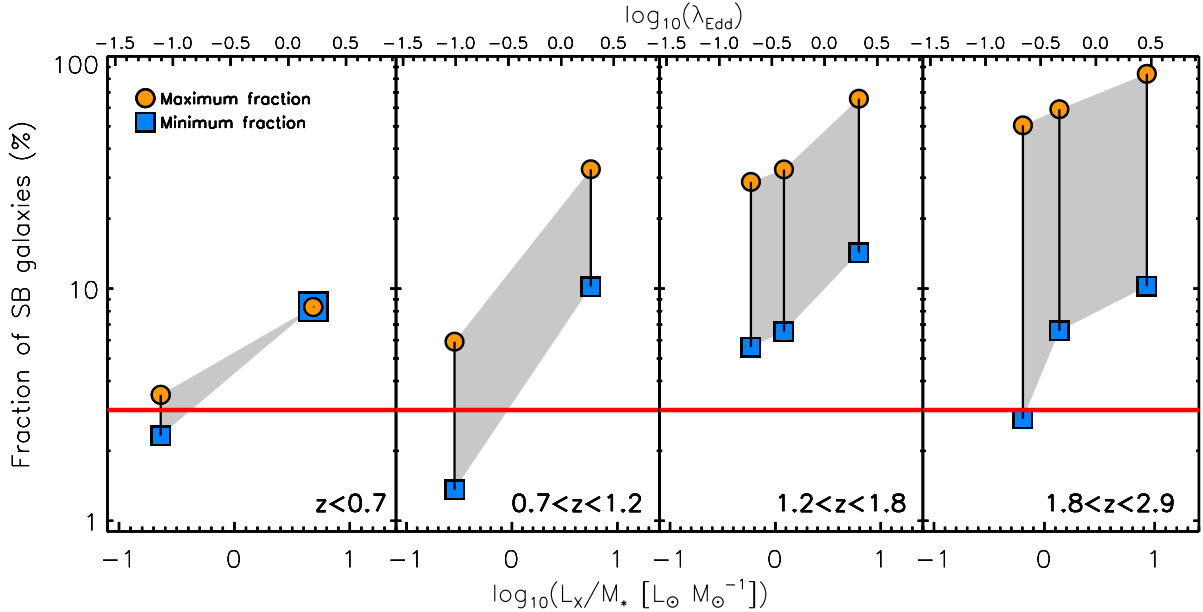}
  \caption{Fraction of starbursting hosts in each of our redshift and specific X-ray luminosity (bottom axis) or \protect \Ledd\ (top axis) bins. The blue squares show the minimum fraction of starburst (i.e., all upper limits measured in the starburst regime are assumed to be non-starburst) and the orange circles show the maximum fraction of starburst (i.e., all upper limits measured in the starburst regime are assumed to be starbursts) in each of our bins. The gray area shows the range of possible values. The red line represents the expected fraction of starburst for a non-AGN star-forming population composed of MS and starburst galaxies only (i.e., 3 per cent; from \protect \citealt{Schreiber2015}). In three of our redshift bins, we find that the minimum fraction of starburst among high \Ledd\ AGNs is significantly higher than that of the general population of star-forming galaxies.}
  \label{fig:sbfrac}
\end{figure*}

\subsection{The fraction of starburst galaxies among AGN hosts}
\label{subsec:fraction} 

The main goal of this study is to investigate how the star-forming properties of AGN host galaxies relate to the specific X-ray luminosity. In \S\ref{subsec:lagnsfr}, we found that there is no correlation between average SFR and specific AGN luminosity, at least up to redshift $\approx3$ and for $0.1< L_{X}/M_{\ast} <10~L_{\odot~}M_{\odot}^{-1}$ (see also Fig.~\ref{fig:sfreddrat}). However, when we instead compare the star-forming properties of X-ray selected AGN host galaxies to a mass-matched sample of MS galaxies, we find that high \Ledd\ AGNs have slightly higher normalised average SFRs compared to low \Ledd\ AGNs (at 3.13 significance for z$\sim$2; see Fig.~\ref{fig:lagnsfr}). This measured difference suggests that the distribution of AGN host SFRs must be different between the lowest and the highest \Ledd\ bin.

To explore this further we attempt to compare the distribution of star-forming properties of AGN hosts across each of our bins, as well as against the distribution of non-AGN star-forming galaxies \citep[i.e.,][]{Sargent2012, Schreiber2015}. Due to the low {\it Herschel} detection rate of individual AGNs (i.e., $\sim$20 per cent of the sample) the faint end of the SFR distribution cannot be probed directly. Instead, we focus on the top end of the SFR distribution by constraining the fraction of our samples that have normalised SFRs a factor five above the MS, which is the mode of the SFR distribution of the so-called ``starburst'' galaxies as defined in \cite{Schreiber2015}. 

When deriving fractions of samples above and below a given threshold, it is important to consider non-detections (in the form of upper-limits) as well as detected sources. To account for {\it Herschel}-undetected AGNs we derive upper-limits on their SFRs by measuring the 3$\sigma$ upper-limits on their {\it Herschel} fluxes. To derive the flux upper-limit, we take three times the standard deviation of a flux distribution given by measuring the flux in 100 positions selected at random around the source (i.e., between twice and 4 times the full width at half maximum of the PSF). We use the \cite{Mullaney2011} star-forming galaxies templates to convert the monochromatic 3$\sigma$ upper-limits at 100~\mum\ for  {\it Herschel}-undetected sources at $z \leq 1$ and at 160~\mum\ for those at $z>1$. Since some of the undetected AGNs in our samples show upper-limits above our starburst threshold (defined above), the maximum starburst fraction is defined by assuming that all these upper-limits represent genuine starburst AGNs. Conversely, the minimum fraction of starburst is defined by assuming that all upper-limits above the starburst threshold are, in fact, non-starburst galaxies. Therefore for each of our redshift and \Ledd\ bins we have a (conservative) range of possible values for the fraction of starburst (i.e., from the minimum to the maximum). Since the $z>2.9$ redshift bin contains only 114 AGNs, of which 91~per cent have only upper-limits on their SFRs, we do not consider this bin for studying the evolution of the fraction of starburst with \Ledd . The results are displayed in Fig.~\ref{fig:sbfrac}, in which we show the range of possible starburst fractions as a function of specific X-ray luminosity (bottom axis in Fig.~\ref{fig:sbfrac}), or \Ledd\ (top axis in Fig.~\ref{fig:sbfrac}) and redshift. For two of our redshift bins (i.e., $z<$0.7 and 0.7$<z<$1.2), we measure an increase of the fraction of starburst, with the minimum fraction in our log(\Ledd )$>0.5$ bin which is higher by $\approx$5 percentage points than the maximum fraction in our $\log($\Ledd $)<0.5$ bin (i.e., $\approx8$~per cent vs. $\approx4$~per cent for $z<$0.7, and $\approx6$~per cent vs. $\approx10$~per cent for 0.7$<z<$1.2). Again, this suggests that the distribution of normalised SFRs changes with \Ledd, such that AGNs of higher \Ledd\ have, on average, marginally higher normalised SFRs. However, for the redshift bins 1.2$<z<$1.8 and 1.8$<z<$2.9, the range of possible starburst fractions is too broad to be able to claim any significant change in the fraction of starbursting hosts as a function of \Ledd .

In Fig.~\ref{fig:sbfrac} we also show the fraction of non-AGN galaxies that we would expect to see in the general star-forming galaxy population. This is calculated from the normalised SFR distribution of \cite{Schreiber2015} and is $\approx 3$~per cent. We use our mass-matched sample to confirm that this fraction is appropriate for our sample (i.e., is not subject to non-linear effects arising from binning in \Ledd). In all of our redshift bins, we find that the {\it minimum} fraction of starburst galaxies in our highest \Ledd\ bins is higher than in the general star-forming galaxy population, with fractions of 8.3~per cent, 10.2~per cent, 14.3~per cent and 10.2~per cent in our $z\sim 0.4, 1, 1.5$ and $2.5$ bins. We calculate the Poisson uncertainty on these percentages, which allows us to measure the significance of the difference between the fraction of starburst galaxies in each of our redshift and highest \Ledd\ bins and the expected 3 per cent, using:

\begin{equation} 
  P(k|\lambda)=\frac{\lambda^k e^{-\lambda}}{k!}
\end{equation}

\noindent where $k$ is the {\it minimum} number of starburst galaxies in each bin (calculated from our minimum fraction) and $\lambda$ is the number that we would expect if 3 per cent of the AGNs in that bin were starbursts. Thus $1-P(k|\lambda)$ gives the probability of measuring the observed fraction by chance if AGNs and star-forming galaxies had the same fraction of starbursts. The highest \Ledd\ bins in our $z\sim 0.4, 1, 1.5$ and $2.4$ bins all have starburst fractions that are higher (at 72.9~per cent, 98.5~per cent, 99.99~per cent and 99.8~per cent confidence, respectively) than the star-forming galaxy population. 


\section{Discussion}
\label{sec:discuss}

In this study, we have explored the star-forming properties of AGNs as a function of their specific X-ray luminosity -- used as a proxy measure of Eddington ratio \Ledd\ -- up to redshift $\approx$3. This work expands upon that of \cite{Azadi2015} by considering AGNs with higher \Ledd\ and at higher redshifts. Consistent with that study, we find that there is no evidence of a strong correlation between average SFR and \Ledd\ up to $z\approx1$ and to log(\Ledd )$\sim-0.5$. However, by including more extreme AGNs in our sample, we can now also report that there is no evidence of enhanced or suppressed star-formation rates among AGNs with log(\Ledd )$>-0.5$ compared to lower \Ledd\ AGNs, at least up to $z\approx 3$.

When we compare the mean SFR of AGN hosts relative to that of mass and redshift-matched samples of non-AGN star-forming galaxies (from the SFR distribution of \citealt{Schreiber2015}) we find that the highest \Ledd\ AGNs (i.e, log(\Ledd)$>0$) in four of our five redshift bins show marginally enhanced SFRs relative to their lower \Ledd\ counterparts, up to redshift $\approx3$. However, this enhancement is only significant at a level of $>3\sigma$ in one of our redshift bins (i.e., $1.8<z<2.9$). We also find that the {\it minimum} fraction of starburst galaxies in the highest \Ledd\ bin at all redshifts is higher than what would be expected if AGN hosts followed the same SFR distribution as the general star-forming population as presented in \cite{Schreiber2015}. These results suggest that the SFR distribution of AGNs changes as a function of specific AGN luminosity and that, at least at high \Ledd\ values, differs from that of normal (i.e., non-AGN) star-forming galaxies. Recently, there has been similar reports of discrepant AGN SFR distributions relative to normal galaxies \citep[e.g.,][]{Shimizu2015, Mullaney2015}. However, these studies found that the mode of the SFR distribution of AGN hosts is roughly a factor of $\sim$2 {\it lower} than that of the MS galaxies, whereas our results indicate a {\it higher} fraction of starbursts among high \Ledd\ AGNs. These results are reconciled by appreciating that \cite{Shimizu2015} and \cite{Mullaney2015} both report that the SFR distribution of AGN hosts is also significantly {\it broader} compared to normal, star-forming galaxies (as defined by e.g., \citealt{Schreiber2015}). This broadened distribution means that there is still an increased likelihood of AGNs residing in strongly starbursting systems, despite the majority of AGNs residing in lower SFR galaxies. Indeed, while we report a higher fraction of starbursts among high \Ledd\ AGNs compared to normal, star-forming galaxies, we also note that roughly 20 per cent of our AGNs in COSMOS are classified as optically quiescent by \cite{Ilbert2013}. While this is only a very approximate estimate of the quiescent fraction, it still underlines the broad diversity of the star-forming properties of AGN hosts. The findings reported here builds upon this by suggesting a link between the SFR distribution and the \Ledd\ of the AGN.

Recently, \cite{Rodighiero2015} and \cite{Delvecchio2015} took the opposite approach to that described here to probe the link between average star-formation and black hole growth rates. Rather than calculate the average star-forming properties of a sample of AGNs, as we do here, they instead calculate the average SMBH growth rate of either a sample of mass-selected \citep{Rodighiero2015} or infrared-selected \citep{Delvecchio2015} galaxies \citep[see also][for earlier studies that take a similar approach]{Rafferty2011, Mullaney2012a, Chen2013}. Both these studies report higher average SMBH growth rates among AGNs with higher SFRs, including those with starburst-like SFRs. Because of the different approaches used, it is difficult to directly interpret our results in terms of these studies. However, there appears to be a certain level of consistency insofar as enhanced SMBH growth rates follows enhanced SFR (or vice-versa). Whether this is a causal relationship, or if there is another more fundamental property linking the two \citep[e.g., the availability of cold gas;][]{Vito2014}, remains unclear.

The higher starburst fraction among high \Ledd\ AGNs compared to that expected from the SFR distribution of normal star-forming galaxies suggests that some aspect of the starbursting process is conducive to triggering some high \Ledd\ AGNs. A number of studies have reported that strongly starbursting systems are triggered by major galaxy mergers \citep[e.g.,][]{Noguchi1988, Hernquist1989, Barnes1991, Duc1997b, Elbaz2003, Barnes2004, Davies2015}. As such, our results could be interpreted that the merger process also plays a role in triggering a proportion of high \Ledd\ AGNs. This is consistent with previous finding that report hightened numbers of powerful AGNs among merging systems \citep[e.g.,][]{Treiste2012, Treiste2013, Ellison2015}. However, it remains unclear whether the merger plays a direct role in triggering powerful AGNs, or whether it is the resultant star-formation that is the more important fueling mechanism \citep[e.g.,][]{Thompson2005, Cuadra2006, Cuadra2008, Vollmer2008}. However, we again stress that our results do not imply that the majority of high \Ledd\ AGNs reside in starbursts (which {\it may} be merger-induced), simply that high \Ledd\ AGNs are slightly more predisposed to starbursting systems compared to the normal galaxy population.

A caveat of using the FIR emission of the host galaxy to derive the SFR is that it measures the integrated SFR on a timescale of several hundreds of million years.  By contrast, it is thought that a typical lifetime of an AGN phase is of the order of $10^7$--$10^{9}$ years \citep[e.g.,][]{Martini2001, Marconi2004}, during which time the accretion rate of the SMBH likely fluctuates by several order of magnitudes on even shorter timescales \citep[e.g.,][]{Alexander2012, Neistein2014, Hickox2014}. Since we use the X-ray emission of the nucleus to derive the instantaneous power of the AGN it may be difficult to interpret the impact that AGNs have on SFRs of host galaxies by directly comparing this to the infrared output of galaxies. However, using hydrodynamical simulations, \cite{Volonteri2015} were able to track the time evolution of the black hole accretion rate versus the SFR for different stages of a major merger (i.e., before, during and after the proper merger), including the rapid variability of AGNs, and averaging the SFR over 100 million years to mimic the FIR estimates. They found that during the merger phase (i.e., enhanced SFR),  SMBHs spend a greater fraction of time accreting at high Eddington ratios compared to during non-merger phases. They also found that high accretion rate phases are often concurrent with a starburst, but that supernova feedback arising from this starburst suppresses AGN activity. In these simulations, therefore, the distribution of SFR, averaged over 100 million years, for high Eddington ratio AGN is broader, and it also peaks at larger SFR than for low Eddington ratio AGN. In such a scenario, it may be more {\it likely} that we find high Eddington ratio AGNs during a merger phase than in a non-merger phase, depending on the relative times spent during those phases. Considering that major mergers are thought to help induce starbursts \citep[e.g.,][]{Noguchi1988, Hernquist1989, Barnes1991, Duc1997b, Elbaz2003, Barnes2004, Davies2015} these results could therefore be considered as being consistent with our findings of a higher fraction of starbursting galaxies among high \Ledd\ AGNs. Whether or not our highest \Ledd\ bins are dominated by galaxies undergoing major mergers is, however, beyond the scope of this paper.

 
\section{Conclusion}
\label{sec:Conclusion}
From the outset, the primary motivation of this study was to determine whether there is any evidence of a suppression of star formation among galaxies hosting the highest Eddington ratio AGNs, which could be interpreted as evidence of so-called ``AGN feedback''. To aim this, we used deep {\it Herschel} data to measure the star-forming properties of a sample of \sampsize\ X-ray selected AGNs in the COSMOS, GOODS-N and GOODS-S fields. Our AGN sample spans $10^{42}< L_{x}<10^{46}$ and $z\lesssim$3 (see \S\ref{sec:dat}). Rather than measuring the Eddington ratio directly, we calculate AGN luminosity relative to the stellar mass of the host galaxy (i.e., the specific $L_{\rm AGN}$); in some respects, this is more pertinent than the actual Eddington ratio, as it provides a measure of the AGN power relative to the gravitational potential of its host. Stacking analyses were employed to account for {\it Herschel}-undetected sources, and broad-band SED fitting was used to account for any AGN contamination when deriving host stellar masses and star-formation rates.

In agreement with previous studies \citep[e.g.,][]{Lutz2010, Shao2010, Mullaney2012a, Rosario2012, Rovilos2012, Harrison2012, Stanley2015}, we found no clear evidence of any strong (anti-)correlation between AGN luminosity and the star-forming properties of their hosts, at least for the luminosity and redshift ranges considered here. When we instead consider specific $L_{\rm AGN}$, however, we find a small increase in the star-formation rates among the highest specific $L_{\rm AGN}$ AGNs; although this increase is only significant at a $>3\sigma$ level in one of our redshift bins. One means of investigating this further will be to derive the distribution of star-forming properties of AGN hosts compared to the non-AGN population. A full exploration of this is beyond the scope of this work, but we do find that the fractions of strongly starbursting are at least a factor of $\sim3$ times larger among our highest specific $L_{\rm AGN}$ AGNs compared to the non-AGN population. From this, we argue that the distribution of SFRs among the AGN population is broader than that observed for MS galaxies, which is consistent with the findings of other recent studies.


\section{acknowledgements}
We thank the anonynous referee for comments that help to improve the quality of the paper. We thank C. Tadhunter and M. Volonteri for the helpful discussions. We also thank M. Volonteri for providing details of her hydrodynamical simulations. The scientific results reported in this article are based on data obtained from the Chandra Data Archive. This research has made use of data from HerMES project (\url{http://hermes.sussex.ac.uk/}). HerMES is a Herschel Key Programme utilising Guaranteed Time from the SPIRE instrument team, ESAC scientists and a mission scientist. The GOODS-{\it Herschel} and the HerMEs data was accessed through the Herschel Database in Marseille (HeDaM - \url {http://hedam.lam.fr}) operated by CeSAM and hosted by the Laboratoire d'Astrophysique de Marseille.

\bibliographystyle{mnras}

\bibliography{biblio}

\end{document}